\documentclass[12pt]{iopart}
\usepackage{iopams}
\usepackage{setstack}
\usepackage[dvips]{graphicx}

\begin{document}

\title[Polarization squeezing]{Polarization squeezing in a 4-level system}

\author{V. Josse\footnote[3]{To whom correspondence should be addressed
(josse@spectro.jussieu.fr)}, A. Dantan, A. Bramati, M. Pinard, E.
Giacobino}

\address{Laboratoire Kastler Brossel, Universit\'{e} Pierre et Marie Curie,
4 place Jussieu, F75252 Paris Cedex 05, France}

\begin{abstract}
We present a theoretical study of an ensemble of X-like 4-level
atoms placed in an optical cavity driven by a linearly polarized
field. We show that the self-rotation (SR) process leads to
polarization switching (PS). Below the PS threshold, both the mean
field mode and the orthogonal vacuum mode are squeezed. We provide
a simple analysis of the phenomena responsible for the squeezing
and trace the origin of vacuum squeezing not to SR, but to crossed
Kerr effect induced by the mean field. Last, we show that this
vacuum squeezing can be interpreted as \textit{polarization
squeezing}.
\end{abstract}




\section{Introduction}

The principal limit in high precision measurements and optics
communication is given by the quantum fluctuations of light. For
several years, in order to beat the standard quantum limit, a
number of methods consisting in generating squeezed states of
light have been developed \cite{davidovich}. In connection with
quantum information technology the quantum features of the
polarization of light has raised a lot of attention. The
generation of polarization squeezing has been achieved
experimentally by mixing an OPO-produced squeezed vacuum with a
coherent field \cite{grangier,polzik}, or more recently by mixing
two independent OPA-originated squeezed beams on a polarizing
beamsplitter \cite{bachor}. Several schemes using Kerr-like media
have also been proposed \cite{chirkin2,boivin, korolkova3}, and
very recently, Matsko et al. proposed to propagate a linearly
polarized field through a self-rotative atomic medium to produce
vacuum squeezing on the orthogonal polarization \cite{matsko}. The
Kerr-like interaction between cold cesium atoms placed in a high
finesse optical cavity and a circularly polarized field has been
studied in our group and a field noise reduction of 40\% has been
obtained \cite{lambrecht,coudreau}. We recently observed
experimental evidence of polarization squeezing when the incoming
polarization is linear \cite{josse}. In this paper, we present a
theoretical investigation of polarization squeezing generated by
an ensemble of X-like 4-level atoms illuminated by a linearly
polarized field. To be as realistic as possible, the experimental
parameters values of Ref \cite{lambrecht,coudreau,josse} are taken
as references. In the first part of the paper, we give a detailed
study of the steady state and show that self-rotation is
responsible for polarization switching and saturation leads to
tristability. We derive simple analytical criteria for the
existence of elliptically polarized solutions and the stability of
the linearly polarized solution. This steady state study is
essential to figure out the interesting working points for
squeezing. In the second part, we focus on the case in which the
polarization remains linear (below the PS threshold) and show that
both the linearly polarized field mode and the orthogonal vacuum
mode are squeezed. Analytical spectra are derived in the low
saturation limit and enable a clear discussion of the physical
effects responsible for polarization squeezing; in particular, we
demonstrate that self-rotation is associated to strong atomic
noise terms preventing vacuum squeezing at low frequency. On the
other hand, saturation accounts for the squeezing on the mean
field and crossed-Kerr effect enables to retrieve vacuum squeezing
at high frequency. The analytical results are compared with a full
quantum calculation. Finally, we derive the Stokes parameters
\cite{chirkin} and relate their fluctuations to those of the
vacuum field. The vacuum squeezing obtained is then equivalent to
the squeezing of one Stokes parameter, the so-called
\textit{polarization squeezing} \cite{korolkova}.

\section{The model}

\begin{figure}[h]
  \centering
  \includegraphics[width=7cm]{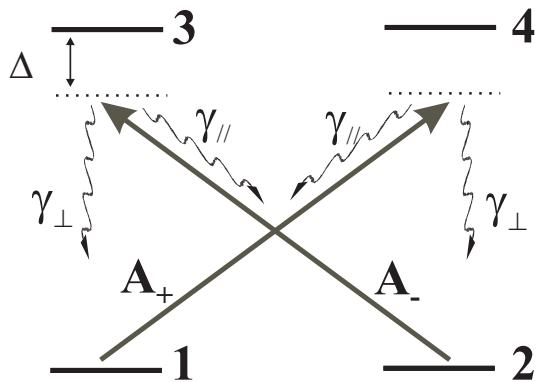}
  \caption{X-like 4-level configuration.}\label{fig1}
\end{figure}

The system considered in this paper is a set of N 4-level cold
atoms interacting in an optical cavity driven by a linearly
polarized field as represented in Fig \ref{fig1}. We denote
$A_\pm$ the slowly-varying envelope operators associated with the
$\sigma_{\pm}$ components of the light \cite{fabre}. They are
defined from the standard linear polarization components

\begin{equation}
A_+=-\frac{A_x-iA_y}{\sqrt{2}}\;\;,\;\;\;A_-=\frac{A_x+iA_y}{\sqrt{2}}
\label{passage}
\end{equation}

The atomic frequencies are both equal to
$\omega_{at}=\omega_{13}=\omega_{24}$. The field frequency is
$\omega$ and the detunings from atomic resonance are equal on both
transitions to $\Delta=\omega_{at}-\omega$. The 4-level system is
described using collective operators for the N atoms of the
ensemble, the optical dipoles being defined in the rotating frame
associated to the laser frequency (e.g.
$\sigma_{14}=\sum\limits_{i=1}^N e^{i\omega t}|1\rangle_i\langle
4|_i$). The coupling constant between the atoms and the field is
defined by $g=\mathcal{E}_0 d/\hbar$, where $d$ is the atomic
dipole and
$\mathcal{E}_0=\sqrt{\hbar\omega/2\epsilon_0\mathcal{S}c}$. With
this definition, the mean square value of the field is expressed
in number of photons per second. As in Fig \ref{fig1}, the
population of level 3 decays with rate $2\gamma_{\perp}$ on level
1 and with rate $2\gamma_{//}$ on level 2, the dipole decay rate
being $\gamma=\gamma_{//}+\gamma_{\perp}$. We consider the case of
saturated optical pumping and neglect the relaxation rate of the
ground states populations. This approximation is well verified for
alkali cold atoms \cite{chu}. With these conventions, the
atom-field hamiltonian is

\begin{equation}
H=\hbar
g[A_+\sigma_{41}+A_+^{\dagger}\sigma_{14}+A_-\sigma_{32}+A_-^{\dagger}\sigma_{23}]
\end{equation}

The atomic evolution is then governed by a set of quantum
Heisenberg-Langevin equations

\begin{eqnarray}
\frac{d\sigma_{14}}{dt} & = & -(\gamma+i\Delta)\sigma_{14}-ig
A_+(\sigma_{11}-\sigma_{44})+F_{14}\\
\frac{d\sigma_{23}}{dt} & = & -(\gamma+i\Delta)\sigma_{23}-ig
A_-(\sigma_{22}-\sigma_{33})+F_{23}\\
\frac{d\sigma_{11}}{dt} & = &
2\gamma_{\perp}\sigma_{33}+2\gamma_{//}\sigma_{44}-ig
(A_+^{\dagger}\sigma_{14}-A_+\sigma_{41})+F_{11}\\
\frac{d\sigma_{22}}{dt} & = &
2\gamma_{//}\sigma_{33}+2\gamma_{\perp}\sigma_{44}-ig
(A_-^{\dagger}\sigma_{23}-A_-\sigma_{32})+F_{22}\\
\frac{d\sigma_{33}}{dt} & = & -2\gamma\sigma_{33}+ig
(A_-^{\dagger}\sigma_{23}-A_-\sigma_{32})+F_{33}\\
\frac{d\sigma_{44}}{dt} & = & -2\gamma\sigma_{44}+ig
(A_+^{\dagger}\sigma_{14}-A_+\sigma_{41})+F_{44}
\end{eqnarray}

Note that we have not reproduced all the atomic equations, but
only those of interest for the following. The Langevin operators
$F_{\mu\nu}$ are $\delta$-correlated and their correlation
functions are calculated via the quantum regression theorem
\cite{cohen}. We consider a ring cavity with $T$ the transmission
of the cavity coupling mirror, $\omega_c$ the cavity resonance
frequency closest to $\omega$ and $\tau$ the cavity round-trip
time. The cavity dephasing is $\Phi_c=(\omega-\omega_c)\tau$. The
incoming quantum fields are $A_{\pm}^{in}$ and the field equations
read

\begin{eqnarray}
\tau\frac{dA_+}{dt} & = &
-(T/2+i\Phi_c)A_+-ig\sigma_{14}+\sqrt{T}A_+^{in}\label{Aplus}\\
\tau\frac{dA_-}{dt} & = &
-(T/2+i\Phi_c)A_--ig\sigma_{23}+\sqrt{T}A_-^{in}\label{Amoins}
\end{eqnarray}

\section{Steady-state}

\subsection{Atomic steady state}
The atomic steady state is readily obtained by setting the time
derivatives to zero and using the fact that a Langevin operator
mean value is zero. Defining saturation parameters $s_{\pm}$ for
both polarizations,

\begin{eqnarray}
s_{\pm}=\frac{2g^2|\langle
A_{\pm}\rangle|^2}{\Delta^2+\gamma^2}=\frac{\Omega_{\pm}^2/2}{\Delta^2+\gamma^2},\label{S}
\end{eqnarray}

the atomic steady state is given by

\begin{eqnarray}
\langle \sigma_{14}\rangle=\frac{-igN\langle
A_+\rangle}{\gamma+i\Delta}\frac{s_-}{s_++s_-}\frac{1}{1+S} \;,
\langle \sigma_{23}\rangle=\frac{-igN\langle A_-\rangle}{\gamma+i\Delta}\frac{s_+}{s_++s_-}\frac{1}{1+S}\\
\langle \sigma_{11}\rangle=
N\frac{s_-}{s_++s_-}\frac{1+s_+/2}{1+S} \;\;\;\;\;\;\;\;\;\;,
\langle \sigma_{22}\rangle=
N\frac{s_+}{s_++s_-}\frac{1+s_-/2}{1+S}\\
\langle \sigma_{33}\rangle=\langle \sigma_{44}\rangle =
\frac{N}{4}\frac{S}{1+S}\;\;\;with:\;\;\;
S=\frac{2s_+s_-}{s_++s_-}
\end{eqnarray}

$\Omega_{\pm}$ are the Rabi frequencies and $S$ is the coupling
saturation parameter which plays a symmetrical role with respect
to both polarization components. For an x-polarized field,
$S=s_+=s_-=s_x/2$ is directly related to the intracavity field
intensity.

\subsection{Polarization switching}\label{ps}
It is well known that such a coupled system may exhibit
polarization switching when driven by a linearly polarized field
\cite{walls,elisabeth}. In fact, the intracavity field intensities
depend on the atomic dephasings $\Phi_{\pm}$ and absorptions
$a_{\pm}$

\begin{eqnarray}
\Phi_{\pm}=2\Phi_0\frac{s_{\mp}}{s_++s_-}\frac{1}{1+S} & \;\;,\;\; & a_{\pm}=2a_0\frac{s_{\mp}}{s_++s_-}\frac{1}{1+S}\\
\Phi_0=\frac{Ng^2\Delta}{2(\Delta^2+\gamma^2)} & \;\;,\;\; &
a_0=\frac{Ng^2\gamma}{2(\Delta^2+\gamma^2)}
\end{eqnarray}

with $\Phi_0$ and $a_0$ the linear dephasing and absorption in the
absence of saturation. These quantities depend in turn on the
intensities to yield a complex coupled system. In order to derive
analytical criteria for polarization switching, we follow the
method given in \cite{elisabeth} and decompose dephasings and
losses into their linear and non-linear parts,

\begin{eqnarray}
\Phi_{\pm} =  \Phi_l\pm\Phi_{SR}\;\;\; & with\;\;\; &
\Phi_l=\frac{\Phi_0}{1+S}\;,\;\; \Phi_{SR}=\Phi_lx_{SR}\label{phil}\\
a_{\pm} = a_l\pm a_{SR}\;\;\; & with\;\;\; &
a_l=\frac{a_0}{1+S}\;,\;\; a_{SR}=a_lx_{SR}\label{al}
\end{eqnarray}

where $\Phi_{SR}$ and $a_{SR}$ are the non-linear circular
birefringence and dichroism, related to the ellipticity $\epsilon$
\cite{huard}

\begin{equation}
x_{SR}=\frac{s_--s_+}{s_++s_-}=-\sin 2\epsilon \label{epsilon}
\end{equation}

Thus, as pointed out in the literature \cite{savage,yashchuk}, the
optical pumping induces non-linear self-rotation (SR) of
elliptically polarized light. It will be shown in the next section
that this effect is responsible for PS in a cavity configuration.
Let us first focus on the solution for the $\sigma_{\pm}$
components. Normalizing all the dephasings and absorptions by
$T/2$ ($\delta_{j}=2\Phi_{j}/T$ and $\alpha_{j}=2a_{j}/T$), Eqs
(\ref{Aplus}),(\ref{Amoins}) read in steady state

\begin{equation}
s_{\pm}=\frac{s_{max}}{(1+\alpha_l\pm
\alpha_{SR})^2+(\delta_l\pm\delta_{SR}-\delta_c)^2} \label{spm}
\end{equation}

with $s_{max}=2/T s_x^{in}$ the maximal intracavity intensity in
the absence of absorption. Replacing (\ref{spm}) in
(\ref{epsilon}), we derive the equation for $x_{SR}$: non zero
solutions correspond to elliptically polarized states. After
straightforward calculations, we obtain

\begin{eqnarray}
x_{SR}=0 \;\;\;\;&or\;\;\;\;& (\alpha_l^2+\delta_l^2)x_{SR}^2=
\delta_l^2+\alpha_l^2-\delta_c^2-1\label{xsr}
\end{eqnarray}

The first trivial solution corresponds to the linearly polarized
field. It follows from the second equation and
(\ref{phil}),(\ref{al}) that elliptically polarized states may
exist as soon as the existence criterion $C_{ex}$ is satisfied

\begin{equation}
C_{ex}=\frac{\delta_0^2+\alpha_0^2}{(1+S)^2}-\delta_c^2-1\geq
0\;\;\;\;\;\;(s_+\neq s_-) \label{critereexistence}
\end{equation}

Note that the absorption brings a positive contribution to the
existence of asymmetrical solutions: this is due to the fact that
non-linear circular dichroism produces "self-elliptization" of the
field. However, this criterion gives no information on the
stability of the solutions. In order to get some physical insight
into this complicated problem it is useful to look at the
evolution of the linearly polarized solution.

\subsection{Interpretation of polarization switching}\label{opo}
In this section, we give a simple interpretation of PS as the
threshold for laser oscillations. Let us consider the linearly
polarized solution along the x axis. The adiabatic elimination of
the atomic variables leads to

\begin{eqnarray}
\frac{1}{\kappa}\frac{dA_y}{dt}=-(1+i\delta_c)A_y+(i\delta_l-\alpha_l)A_y-
(\delta_{SR}+i\alpha_{SR})A_x+\frac{2}{\sqrt{T}}A_y^{in}\label{equationAy2}
\end{eqnarray}

where $\kappa=T/2\tau$ is the intracavity field decay rate. In
(\ref{equationAy2}) all terms have zero mean value and are of
order 1 in fluctuations ($\langle A_x\rangle\neq 0$). Using
$x_{SR}A_x=i(A_y-A_x^2/|A_x|^2A_y^{\dagger})$, one obtains

\begin{eqnarray}
\frac{1}{\kappa}\frac{dA_y}{dt}=-(1+i\delta_c)A_y+(i\delta_l-\alpha_l)\frac{A_x^2}{|A_x|^2}A_y^{\dagger}
+\frac{2}{\sqrt{T}}A_y^{in} \label{equationAy3}
\end{eqnarray}

Owing to SR the fluctuations of the orthogonal mode undergo a
phase dependent gain. A similar equation has already been derived
in previous theoretical works in a single pass scheme
\cite{matsko}. In our configuration the presence of the cavity
will lead to oscillations of this mode as soon as the phase
sensitive gain is larger than the losses. This condition may be
expressed as follows

\begin{equation}
C_{PS}=\frac{\delta_0^2+\alpha_0^2}{(1+s_x/2)^2}-\delta_c^2-1\geq
0 \label{criterestabilite}
\end{equation}

Obviously, the linearly polarized solution is not stable when
$C_{PS}\geq 0$. However, the adiabatical elimination of the atomic
variables does not \textit{a priori} take all causes for
instability into account. Yet, we checked that this threshold
analysis was consistent with a numerical calculation of the
atom-field stability matrix. In the following we use $C_{PS}$ as a
stability criterion for the linearly polarized solution.
Nevertheless, it does not yield information on the stability of
the elliptically polarized solutions, which has been evaluated numerically.\\
Besides, the ability of a system to produce squeezing being
closely related to its static properties, the fluctuations of the
vacuum field are expected to be strongly modified in the vicinity
of the PS threshold. Since Eq (\ref{equationAy3}) is similar to
that of a degenerate optical parametric oscillator (OPO) below the
threshold \cite{gardiner}, perfect squeezing could be obtained via
SR. However, the atomic noise is not included in
(\ref{equationAy3}) and is to be carefully evaluated.

\subsection{Optical pumping regime}\label{opticalpumping}

\begin{figure}
  \centering
  \includegraphics[width=10cm]{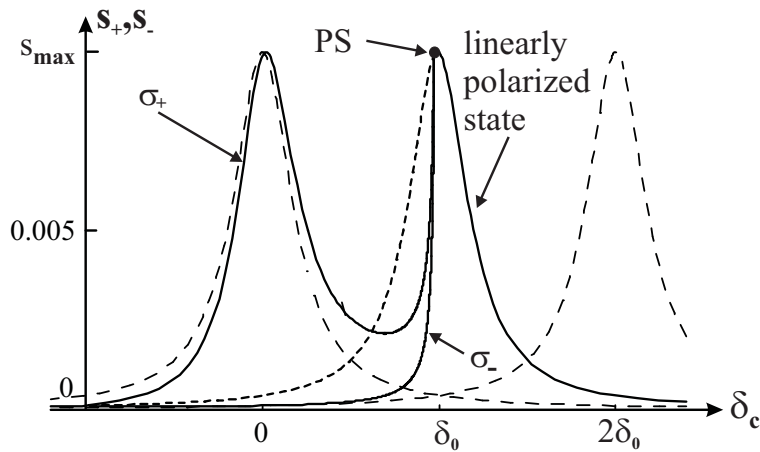}
  \caption{Resonance curves in the optical pumping regime.
  The parameters are $\delta_0=5$, $s_{max}=0.01$. The cavity dephasing
  corresponding to PS is $\delta_ {PS}=4.85$, close to $\delta_0$
  as given by the PS criterion. When the cavity detuning is scanned from the
  right, the linear solution is stable until $\delta_c=\delta_{PS}$
  and unstable afterwards. Then
  the elliptically polarized solutions, $s_+$ and $s_-$,
  become stable for $|\delta_c|\leq\delta_{PS}$.
  We plot also the resonance peaks (dashed line) for the cavity with $0$ or $N$ atoms, in the
  absence of SR phenomenon.}\label{fig2}
\end{figure}

PS is caused by a competition between the two $\sigma_{\pm}$
optical pumping processes. We can understand the main features of
this effect by restraining ourselves to the case where absorption
and saturation are negligible: $\Delta\gg\gamma$ and $s_x\ll 1$.
Neglecting the excited state populations, the optical pumping
equations for the ground state populations are

\begin{eqnarray}
\left(\frac{d\sigma_{11}}{dt}\right)_{pumping}
=-\gamma_{\perp}s_+\sigma_{11}+\gamma_{\perp}s_-\sigma_{22}\label{op1}\\
\left(\frac{d\sigma_{22}}{dt}\right)_{pumping}
=-\gamma_{\perp}s_-\sigma_{22}+\gamma_{\perp}s_+\sigma_{11}\label{op2}
\end{eqnarray}

so that the $\sigma_+$ component tends to pump the atoms into
level 2, the $\sigma_-$ into 1, and, in steady state,
$\sigma_{11}\propto s_-$ and $\sigma_{22}\propto s_+$. The
circular birefringence $\delta_{SR}$ is proportional to the ground
state population difference, and consequently, to the intensity
difference $s_+-s_-$ [see (\ref{epsilon})]. This simple analysis
allows for relating self-rotation to competitive optical
pumping and will help us interpret the resonance curves.\\
Under the previous conditions both criteria
(\ref{critereexistence}) and (\ref{criterestabilite}) are
equivalent and it follows that the linearly polarized solution
bifurcates into an elliptically polarized state for
$|\delta_c|\leq\delta_{PS}=\sqrt{\delta_0^2-1}$. Consequently, PS
is observed as soon as the linear dephasing is greater than half
the cavity bandwidth ($\delta_0\geq 1$). This represents an easily
accessible condition from an experimental point of view : in our
cesium experiment using a magneto-optical trap
\cite{lambrecht,coudreau}, the number of atoms interacting with
the light is $N\simeq 7\;10^6$. To find realistic experimental
parameters, we assimilate each one of our X-model transitions to
the transition $6S_{1/2}\;(F=4)-6P_{3/2}\;(F=5)$ of the $D_2$ line
of $^{133}Cs$, for which $\gamma/2\pi=2.6$ MHz. The square of the
coupling constant $g$ is proportional to the ratio of the
diffusion section at resonance to the transversal surface $S=0.1$
mm$^2$ of the beam, $g^2=3\gamma\lambda^2/4\pi S=4.24$ Hz. The
cavity transmission is 10\%. To obtain a sufficiently high
non-linearity, keeping the absorption low, a good detuning is
$\Delta\simeq 20\gamma$, so that an approximate value for the
linear detuning is $\delta_0\simeq 5$. Note that the saturation
parameters of (\ref{S}) are simply

\begin{equation}
s_{\pm}=\frac{\gamma^2}{\gamma^2+\Delta^2}\frac{I_{\pm}}{I_{sat}}
\end{equation}

The saturation intensity being
$I_{sat}=\epsilon_0c\gamma^2\hbar^2/d^2=1.05$ mW/cm$^2$ \cite{lambrecht},
typical values for $s_{\pm}$ are 0.1-1.\\

In Fig \ref{fig2} are represented the admissible intensities for
the $\sigma_+$ and $\sigma_-$ components versus the cavity
detuning for typical experimental values of the parameters. The
peak centered on $\delta_c=\delta_0$ corresponds to the
symmetrical solution. When the cavity is scanned from right to
left, the linearly polarized field ($s_+=s_-$) intensity increases
until the PS threshold is reached
($\delta_c=\delta_{PS}\simeq\delta_0$). Then one elliptically
polarized state becomes stable. The predominant circular
component, say $\sigma_+$, creates, via the optical pumping
process (\ref{op1}-\ref{op2}), a positive orientation of the
medium $\sigma_{11}\simeq 0$, $\sigma_{22}\simeq N$. Since the
atomic dephasing decreases to zero for the $\sigma_+$ component
($\delta_+\simeq 0$), as if it were propagating in an empty
cavity. Hence, the solution draws close to the zero-dephasing
peak, that is, close to resonance in the range
$|\delta_c|\leq\delta_{PS}$. On the other hand the $\sigma_-$
component "sees" all the atoms ($\delta_-\simeq 2\delta_0$) and
breaks down to fit the peak centered on $\delta_c=2\delta_0$,
which is far from resonance. In order to illustrate this
interpretation of the resonance curves, the two Airy peaks
centered on $\delta_c=0$ and $\delta_c=2\delta_0$ are represented
in Fig \ref{fig2}. As the cavity detuning is decreased both
asymmetrical solutions reunite when the criterion
(\ref{criterestabilite}) is no longer satisfied and the linear
solution becomes stable again. These simple interpretations will
help us understand the much more complex general case, when
absorption and saturation come into play.\\
As discussed in Ref \cite{elisabeth}, taking into account the
ground state relaxation rate $\gamma_0$ yields tristability in the
unsaturated optical pumping regime
($\gamma_{\perp}s_{\pm}\ll\gamma_0$). We will now show that the
optical saturation also leads to tristability.

\subsection{Tristability}\label{kerrstat}

\begin{figure}
  \flushright
  \includegraphics[width=13cm]{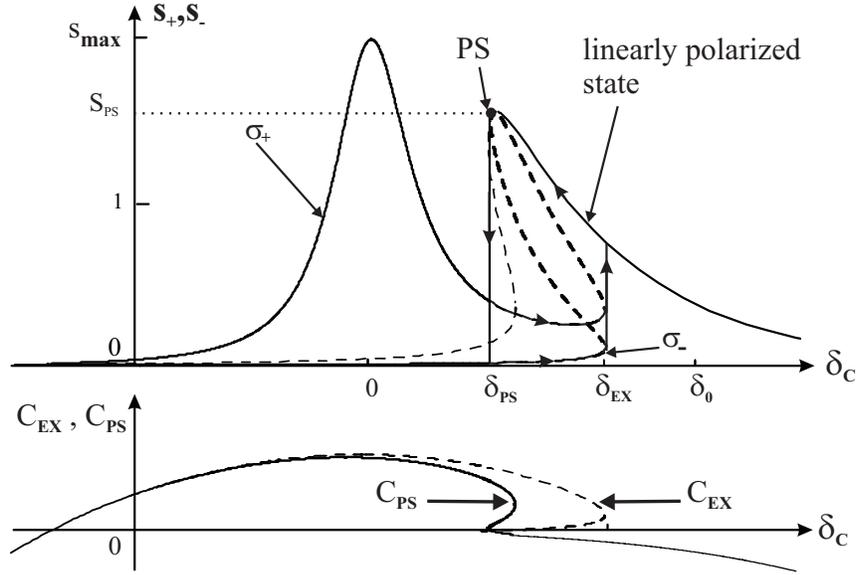}
  \caption{Upper plot: resonance curves for $\delta_0=7$,
  $\Delta=20\gamma$ and $s_{max}=2$. The linear absorption is $\alpha_0=0.35$. Dashed
  parts indicate unstable solutions. The switching occurs for
  $\delta_{PS}=2.6$ and $S_{PS}=1.5$. $C_{ex}=0$ for $\delta_{ex}=5.6$, so that
  the tristability range is $2.6\leq\delta_c\leq5.6$.
  The arrows on the hysteresis cycle correspond to increasing and decreasing cavity
  detuning scan.
  Below are plotted the two criteria: $C_{PS}$ giving the stability
  of the linear solution (plain) and
  $C_{ex}$ giving the existence of asymmetrical solutions (dashed).}\label{fig3}
\end{figure}

It is well-known that saturation may induce multistability for the
linearly polarized field \cite{gibbs} in our configuration and
substantially modify the steady state. When the non-linearity is
sufficient, there may be three possible values for the x-polarized
field intensity. Therefore, saturation is an additional cause of
instability for the symmetrical solution. In Fig \ref{fig3}, we
plotted the same curves as in Fig \ref{fig2}, but for higher
values of $s_{max}$ and $\delta_0$. As expected, the linearly
polarized state solution is distorted as a consequence of the
non-linear effect. The effect of absorption is also clear: whereas
the symmetrical peak height is reduced, the $\sigma_+$ dominant
peak height is not. Indeed,
the $\sigma_+$ component "sees" no atoms after the switching.\\
Besides, the system now exhibits tristability for a certain range
of the cavity detuning. As mentioned previously the existence of
asymmetrical solutions is related to the positivity of $C_{ex}$,
whereas the stability of the symmetrical solution is given by
$C_{PS}$. For instance, on Fig \ref{fig3}, $C_{ex}=0$ for
$\delta_c=\delta_{ex}=5.6$ and the threshold $C_{PS}=0$ is reached
for a dephasing $\delta_c=\delta_{PS}=2.6$. Thus, in the range
$\delta_{PS}\leq\delta_c\leq \delta_{ex}$, two different sets of
asymmetrical solutions exist, in addition to the linear
polarization state. This phenomenon is due to the saturation
experienced by the $\sigma_+$ and $\sigma_-$ components. As
expected, we checked that only the lower branch of each
asymmetrical curve is stable, leading to tristability for the
polarization state : linear, $\sigma_+$-dominant or
$\sigma_-$-dominant. The system switches for a different value of
the cavity detuning if the cavity is scanned from left to right,
or from right to left (see Fig \ref{fig3}). Hence, unlike the
unsaturated case, saturation induces a multistable behavior and a
hysteresis cycle now appears in the resonance curve.\\

This brief study of the resonance curve for typical parameters
leads to an essential observation: the lower branch of the
bistability curve for the linear polarized field is not stable. We
may wonder if there is a domain of the parameter space for which
it is not the case. Since the quantum fluctuations are expected to
be most reduced in the vicinity of the lower turning point
\cite{reynaud,hilico}, the answer is of crucial importance for
squeezing and will be treated in the next section.

\subsection{Competition between SR and
saturation}\label{competition}

\begin{figure}
  \centering
  \includegraphics[width=10cm]{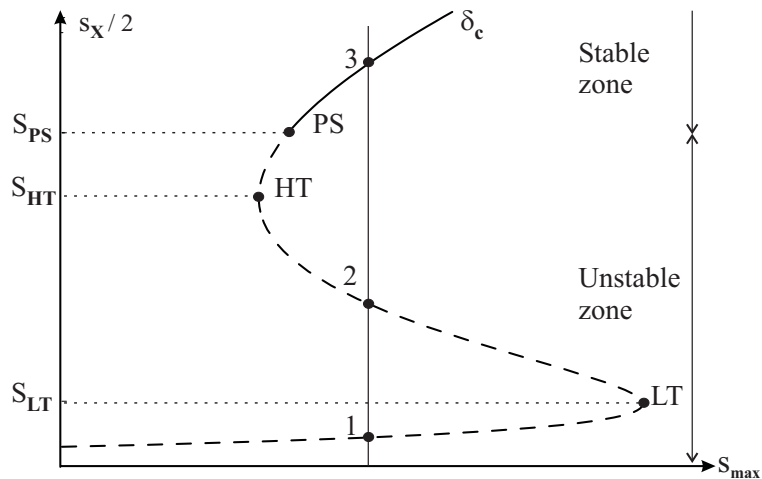}
  \caption{Bistability curve: linearly polarized field intensity $S=s_x/2$ as a function of
  $s_{max}$. PS is the switching threshold, $HT$ and $LT$ the higher and lower turning
  points. For certain incident intensities $s_{max}$ three
  solutions (1,2,3) exist for the intracavity intensity $s_x/2$, of which
  only one (3) is stable, instead of the usual two (1,3) in the
  absence of PS phenomenon.
  }\label{fig4}
\end{figure}

\begin{figure}
  \centering
  \includegraphics[width=11cm]{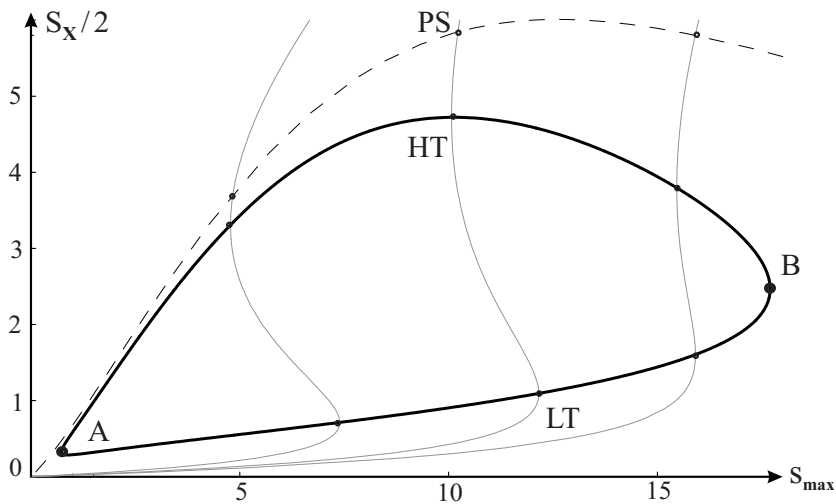}
  \caption{Bistability curves $S=s_x/2$ as a function of $s_{max}$ for field $A_x$.
  The three S-shaped curves correspond to different values of the cavity dephasing
  ($\delta_c=-0.25,0.3,1.1$ from left to right),
  $\delta_0$ and $\Delta$ having the same value as in Fig \ref{fig3}.
  The AB segments represent the $HT$ (higher) and $LT$ (lower) curves.
  The dashed curve is the ensemble of the intensities $S_{PS}$ for which
  polarization switching occurs. The system exhibits bistability for $s_A\leq s_{max}\leq s_B$.}\label{fig5}
\end{figure}

To complete the analysis of the steady state, we would like to
emphasize that, when the cavity is scanned, PS always happens
before reaching the higher turning point of the bistability curve.
In order to get some insight into this complicated problem, it is
worth looking at Fig \ref{fig4}. We plotted the typical S-shaped
variation of the linearly polarized field intensity $s_x$ versus
the incoming intensity $s_{max}$, for a fixed value of the cavity
detuning $\delta_c$. We choose the parameters so that there is
bistability for this state of polarization and report the position
of the lower ($LT$) and the higher ($HT$) turning points. In the
absence of the PS phenomenon, the solutions between $HT$ and $LT$
are unstable (like 2 on Fig \ref{fig4}), whereas solutions on the
lower (1) and higher (3) branches are stable. However the
stability of the linear polarization is modified by the PS
effects. To a fixed value of the dephasing corresponds the PS
intensity $S_{PS}$ cancelling $C_{PS}$ in
(\ref{criterestabilite}); if $S=s_x/2\leq S_{PS}(\delta_c)$, then
the linear polarization is unstable. Hence, if
$S_{PS}(\delta_c)\geq S_{HT}$ is satisfied in the whole parameter
space, then PS occurs before reaching $HT$, and, consequently, the
lower branch is never stable.\\
This general feature is shown on Fig \ref{fig5}, in which we
represented different bistability curves as in Fig \ref{fig4}. The
upper branch of AB is the $HT$ curve (the ensemble of the higher
turning points when $\delta_c$ is varied), the lower branch is the
$LT$ curve. The dashed curve shows the ensemble of the intensities
$S_{PS}$ for which the polarization switches. This curve is always
above the $HT$ curve, confirming that the linear polarization
always becomes unstable on account of PS first. What is more, we
see that PS is closer to $HT$ for low values of $s_x$. We thus
expect this situation to be the most favorable to achieve
squeezing via optical bistability. We checked that varying the
parameters $\delta_0$ and $\Delta$ does not change the conclusion.

To conclude this section, we would like to point out that
bistability, as well as PS, may disappear when the saturation is
too high, as can be seen from Fig \ref{fig4}. However, we will
focus on the low saturation case in the large detuning limit which
is the most favorable case for squeezing, and provides analytical
results, as well as a clear physical understanding.

\section{Mean field fluctuations}

Since we are interested in the quantum fluctuations, we linearize
the quantum operators around their steady state values following
the standard linear input-output method \cite{reynaud}. The
elliptically polarized solutions are not of great interest for
squeezing since the predominant circular component sees no atom
and the other has negligible intensity. Therefore, in all the
following, we focus on the linearly polarized state and study how
both the mean field $A_x$ and the orthogonal vacuum field $A_y$
may be squeezed. We have calculated the outgoing fields noise
spectra via a full quantum treatment (see e.g. \cite{hilico})
involving the four-level system. The outgoing fields are
standardly defined from the input-output relation \cite{reynaud}:

\begin{equation}
A_{x,\;y}^{out}=\sqrt{T}A_{x,\;y}-A_{x,\;y}^{in}
\end{equation}

Yet, to provide clear interpretations as well as analytical
results, we derive simplified equations, first for
the mean field mode $A_x$, then for the vacuum mode $A_y$.\\
A similar equation to (\ref{equationAy2}) can be derived for the
field $A_x$ with a term arising from SR in $\delta_{SR}A_y$. In
the linearization, this product of zero mean value operators
vanishes, so that we only have to take saturation into account to
derive the spectra of $A_x$. Field squeezing owing to optical
bistability has been widely studied \cite{hilico,lugiato,reid} and
is known to occur on a frequency range given by the cavity
bandwidth $\kappa$. The most favorable configuration is the bad
cavity limit: $\kappa$ is greater than $\gamma$ (in our
experiment, $\kappa\simeq 2\gamma$). In the large detuning limit,
$\Delta\gg g |A_x|\gg\gamma$, the equation for $A_x$ reads at
order 3 in $\gamma/\Delta$,

\begin{equation}
\frac{1}{\kappa}\frac{d}{dt}\delta
A_x=-(1+i\delta_c-i\delta_0)\delta
A_x+i\delta_0\frac{s_x}{2}\left[2\delta
A_x+\frac{A_x^2}{|A_x|^2}\delta
A_x^{\dagger}\right]+\frac{2}{\sqrt{T}}A_x^{in} \label{Ax1}
\end{equation}

where $A_x$ is short for $\langle A_x\rangle$. This simplified
equation yields the classical Kerr terms in $A_x^2\delta
A_x^{\dagger}$ producing squeezing. Note that absorption,
dispersion and the associated atomic noise are not included in
(\ref{Ax1}). The spectra taking absorption and dispersion into
account can be easily derived and are shown on Fig \ref{fig7}. The
associated susceptibility and correlation matrices,
$[\chi]_{Kerr}$ and $[\sigma]_{Kerr}$, of the linear input-output
theory are reproduced in Appendix, and the comparison with a Kerr
medium is discussed in \cite{hilico}. The situation is more
complex for the orthogonal mode on account of SR.

\section{Vacuum fluctuations}

As mentioned in Sec \ref{opo}, SR seems to be a very promising
candidate for generating vacuum squeezing. However, a careful
analysis of the atomic noise, which cannot be neglected, is
necessary in the squeezing calculations. In the optical pumping
regime the circular birefringence, $\delta_{SR}=-2\delta_0J_z/N$,
is proportional to the ground state population difference
$J_z=(\sigma_{22}-\sigma_{11})/2$ (see Sec \ref{opticalpumping}).
The SR effect is thus closely related to the fluctuations of
$J_z$, and consequently to the fluctuations of $A_y$ via the
coupling term in $A_x\delta_{SR}$ [Eq (\ref{equationAy2})].
Therefore, we derive general equations for $\delta A_y$ and
$\delta J_z$ in the Fourier domain, and examine their low and high
frequency limits. For the sake of simplicity, absorption and
linear dispersion, again, are not shown; however, the additional
terms are included in the Appendix. As previously we place
ourselves in the large detuning limit with $s_x\ll 1$ and obtain,
discarding terms of order greater than $(\gamma/\Delta)^3$,

\begin{eqnarray}
\nonumber -i(\omega/\kappa)\delta A_y & = &
-(1+i\delta_c-i\delta_0)\delta
A_y-i\delta_0\frac{s_x}{4}\left[3\delta
A_y-\frac{A_x^2}{|A_x|^2}\delta
A_y^{\dagger}\right]\\
& & +\beta(\omega)\frac{2\delta_0}{N}A_x\delta
J_z+\frac{2}{\sqrt{T}}\delta A_y^{in}+F_{A_y}\label{Ay1}\\
-i\omega\delta J_z & = & -\gamma_p\alpha(\omega)\left[\delta
J_z-\lambda(\omega)\left(1-\frac{s_x}{2}\right)\frac{N}{2}\frac{\delta
S_z}{|A_x|^2}\right]+F_z\label{Jz1}
\end{eqnarray}

where $S_z=i(A_xA_y^{\dagger}-A_x^*A_y)$ is the usual Stokes
parameter (see Sec \ref{polarizationsqueezing}) and

\begin{equation}
\alpha(\omega)=\left[1-\frac{i\omega}{4\gamma_{\perp}}\right]\frac{\beta(\omega)}{\lambda(\omega)}
\;\;,\;\;\;\beta(\omega)=1-\frac{s_x}{4\lambda(\omega)}\;\;,\;\;\;
\lambda(\omega)=\frac{2\gamma-i\omega}{2(\gamma-i\omega)}
\end{equation}

with $F_{A_y}$ and $F_z$ the Langevin operators associated to
$A_y$ and $J_z$ after the adiabatical eliminations,
$\gamma_p=\gamma_{\perp}s_x$ the optical pumping rate. In
(\ref{Ay1}), the last term of the first line is a crossed Kerr
term and clearly contributes to squeezing. We also see that the
coupling with $J_z$ is strongly frequency-dependant and requires a
careful investigation. In the next sections, we discuss the low
and high frequency limits to further simplify the previous
equations and give simple interpretations for the squeezing.

\subsection{Low frequency: SR effect}\label{faradaysection}

\begin{figure}
  \flushright
  \includegraphics[width=14cm]{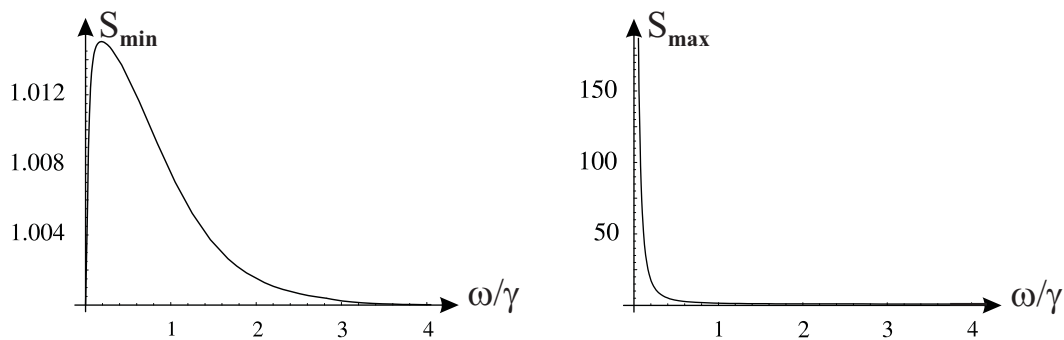}
  \caption{Minimal (left) and maximal (right) outgoing field spectra obtained by considering only SR effects.
  Parameters: $\delta_0=5$, $\Delta=40\gamma$, $\gamma/\gamma_{\perp}=3$,
  $\kappa=2\gamma$, $s_x=0.1$, $\delta_c=3$.}\label{fig6}
\end{figure}

Again, to stress the effects on the fluctuations only due to SR,
we neglect the terms in $s_x$ responsible for the Kerr effect,
which will be studied in the next section, and place ourselves in
the optical pumping regime, keeping only terms of order 1 in
$\gamma/\Delta$. Note that this approximation consists in
adiabatically eliminating the optical dipoles and neglecting the
excited state populations and thus limits the analysis to the
range of frequencies $\omega\ll\gamma,\gamma_{\perp}$. Under these
conditions, one has $\alpha(\omega)\simeq 1$, $\beta(\omega)\simeq
1$, $\lambda(\omega)\simeq 1$ and Eq (\ref{Jz1}) reduces to the
linearized optical pumping equation

\begin{equation}
-i\omega\delta J_z = -\gamma_p\left[\delta
J_z-\frac{N}{2}\frac{\delta S_z}{|A_x|^2}\right]+F_z \label{Jz2}
\end{equation}

It is clear that the fluctuations of $J_z$ are governed by the
time constant $\gamma_p$, consistently with the optical pumping
approximation $\gamma_p\ll\gamma$. SR is effective only at low
frequency. Plugging (\ref{Jz2}) back into (\ref{Ay1}), one gets

\begin{eqnarray}
\nonumber (1+i\delta_c-i\omega/\kappa)\delta A_y & = &
i\delta_0\left[1-\frac{\gamma_p}{\gamma_p-i\omega}\right]\delta
A_y\\& &
+i\delta_0\frac{\gamma_p}{\gamma_p-i\omega}\frac{A_x^2}{|A_x|^2}\delta
A_y^{\dagger}+\frac{2}{\sqrt{T}}\delta A_y^{in}+\tilde{F}_{A_y}
\label{Ay2}
\end{eqnarray}

The SR term comes with an amplitude $\delta_0$
($\propto\gamma/\Delta$) around zero frequency, which is much
greater than the usual third order saturation non-linearity. Very
good squeezing could be expected if it were not for the noise
coming from the atoms $\tilde{F}_{A_y}$, which we now study. The
fluctuation operator arising from atomic and field fluctuations
reads

\begin{equation}
\tilde{F}_{A_y}=\frac{2\delta_0A_x}{N}\frac{F_z}{\gamma_p-i\omega}+F_{A_y}
\end{equation}

The second term $F_{A_y}$ is responsible for the noise due to
absorption. The first term includes the optical pumping noise. One
calculates the correlation function of $F_z$ via the quantum
regression theorem \cite{cohen}

\begin{equation}
\langle
F_z(\omega)F_z(\omega')\rangle=2\pi\delta(\omega+\omega')N\gamma_p/2
\end{equation}

so that

\begin{eqnarray}
\langle
\tilde{F}_{A_y}(\omega)\tilde{F}_{A_y}^{\dagger}(\omega')\rangle&\simeq&
\frac{4\delta_0^2A_x^2}{N^2}\frac{1}{\gamma_p^2+\omega^2}\langle
F_z(\omega)F_z(\omega')\rangle\\
&=&2\pi\delta(\omega+\omega')\frac{C}{4}\frac{\gamma
}{\gamma_{\perp}}\frac{\gamma_p^2}{\gamma_p^2+\omega^2}
\end{eqnarray}

in which we introduced $C=g^2N/T\gamma$ the cooperativity
parameter quantifying the strength of the atom-field coupling via
the cavity ($C\sim100$ in our Cs experiment). For
$\omega\ll\gamma_p$ the noise is thus much more important than the
losses due to absorption and therefore has a dramatic influence on
the squeezing that could have been produced by the SR term.
Following the method given in \cite{hilico}, we derive the
susceptibility and correlation matrices which are given in
Appendix. We can then calculate the outgoing vacuum field spectrum
for all the quadratures. Minimal and maximal spectra are plotted
on Fig \ref{fig6} in the "close-to-bad" cavity limit
($\kappa=2\gamma$) corresponding to our experimental
configuration. Whereas the first is close to the shot-noise level,
the second is extremely noisy. In the good cavity limit, the noise
is even more important. The conclusion is that the optical pumping
process adds too much noise at zero frequency for SR to generate
vacuum squeezing. However, this low frequency noise does not
prevent squeezing at higher frequencies.

\subsection{High frequency limit: crossed Kerr effect}
If one repeats the previous calculation keeping the first order
saturation terms in $s_x$ and considers frequencies
$\omega\gg\gamma$, one finds

\begin{equation}
\delta J_z(\omega)=\frac{N}{2}\frac{s_x}{4}\frac{\delta
S_z(\omega)}{|A_x|^2} \label{Jzkerr}
\end{equation}

This is not surprising, since the evolution times considered are
small with respect to the atomic relaxation time. The system
behaves as if $\sigma_+$ and $\sigma_-$ were independent. In fact,
let us consider two independent two-level systems, 1-4 and 2-3,
each with $N/2$ atoms. In the large detuning limit, one has
$\sigma_{44}=\sigma_{11}s_+/2$ and $\sigma_{33}=\sigma_{22}s_-/2$,
and the atomic fluctuations follow the field fluctuations
\cite{hilico} (still at order 3 in $(\gamma/\Delta)$)

\begin{equation}
\delta \sigma_{11}(\omega)=-\delta
\sigma_{44}(\omega)=-\frac{N}{2}\frac{\delta
s_+}{2}\;\;,\;\;\;\delta \sigma_{22}(\omega)=-\delta
\sigma_{33}(\omega)=-\frac{N}{2}\frac{\delta s_-}{2}
\end{equation}

so that, using $|A_x|^2\delta(s_+-s_-)=s_x\delta S_z$, we retrieve
(\ref{Jzkerr}). This equation shows that the fluctuations of $J_z$
are only caused by saturation and their contribution adds to the
crossed Kerr terms already mentioned in (\ref{Ay1}) to retrieve a
similar "Kerr" equation for $A_y$ to that of $A_x$ at high
frequency

\begin{eqnarray}
\nonumber -i(\omega/\kappa)\delta A_y & = &-(1+i\delta_c-i\delta_0)\delta A_y\\
& &-i\delta_0\frac{s_x}{2}\left[2\delta
A_y-\frac{A_x^2}{|A_x|^2}\delta
A^{\dagger}_y\right]+\frac{2}{\sqrt{T}}\delta A_y^{in}+F_{A_y}
\label{Aykerr}
\end{eqnarray}

This high frequency behavior is thus characterized by the same
Kerr-induced optimal squeezing on both polarization modes,
consistently with the previous analysis for two independent
two-level systems. More precisely, the optimal squeezing spectra
are the same for each mode, but involve orthogonal quadratures
[because of the sign difference in the Kerr terms between
(\ref{Ax1}) and (\ref{Aykerr})]. We now plot the outgoing fields
$A_{x,\;y}^{out}$ squeezing spectra and discuss the squeezing
optimization.

\subsection{Squeezing spectra}
\begin{figure}
  \centering
  \includegraphics[width=10cm]{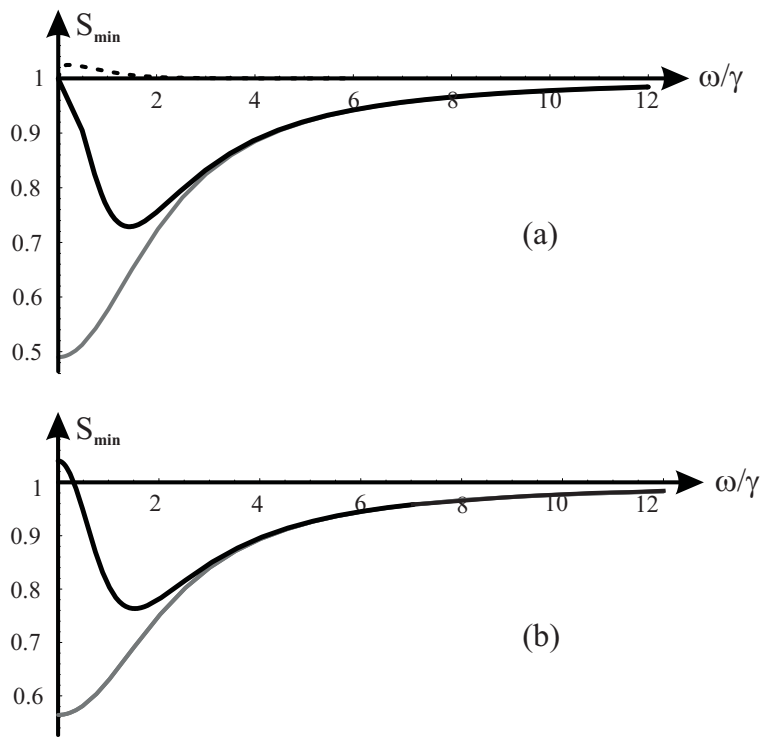}
  \caption{(a) Analytical minimal spectra for $A_y^{out}$ given by the SR effect (dashed),
   by the Kerr effect (light) and by both effects
  (dark).
  (b) Exact spectra for the mean field mode $A_x^{out}$ (light) and the
  orthogonal vacuum mode $A_y^{out}$ (dark). Parameters values:
  $\delta_0=5$, $\Delta=20\gamma$, $\gamma/\gamma_{\perp}=3$,
  $\kappa=2\gamma$, $s_{max}=0.1$, $\delta_c=4.6$.}\label{fig7}
\end{figure}

To derive spectra for the whole frequency range, we combine both
effects by adding the matrices obtained in the two asymptotical
regimes studied previously. We write the complete susceptibility
and correlation matrix under the form
$[\chi(\omega)]_y=[\chi(\omega)]_{Kerr}+[\chi(\omega)]_{SR}$ and
$[\sigma(\omega)]_y=[\sigma(\omega)]_{Kerr}+[\sigma(\omega)]_{SR}$,
where the Kerr matrices are those obtained in the high frequency
limit, and the SR matrices those obtained at low frequency (see
Appendix for analytical expressions). This approximation is good
since Kerr effect is negligible at low frequency compared to SR,
while SR breaks down at high frequency.\\
In Fig \ref{fig7}, typical spectra for a working point close to
the PS threshold are represented. The parameters are chosen to be
as close to the experimental situation as possible \cite{josse}.
We compared these approximate spectra (a) with those obtained with
a full 4-level calculations based on the linear input-output
theory (b). The analytical spectra combining Kerr and SR effects
show indeed an excellent agreement with the exact calculations, as
long as the saturation is low. As shown previously, the SR
spectrum is close to the shot-noise level. The Kerr spectrum is
accurate for $A_y$ only for $\omega\gg\gamma$ and extends on a
range of several $\kappa$ as expected \cite{hilico}. The combined
spectrum (a) confirms that SR destroys completely the squeezing at
low frequency and reproduces well the exact behavior (b). The best
squeezing for $A_y$, obtained at intermediate frequencies, is
about 25\%.\\
Note that the Kerr spectrum in (a) is also valid for field $A_x$
for all frequencies and 45\% of squeezing is obtained at zero
frequency. The situation for the mean field $A_x$ is identical to
that of a circularly polarized field with intensity $s_x/2$
interacting with $N/2$ two-level atoms as in \cite{hilico}, for
which the Kerr spectrum shows good agreement with the exact
spectrum.

\begin{figure}
  \centering
  \includegraphics[width=10cm]{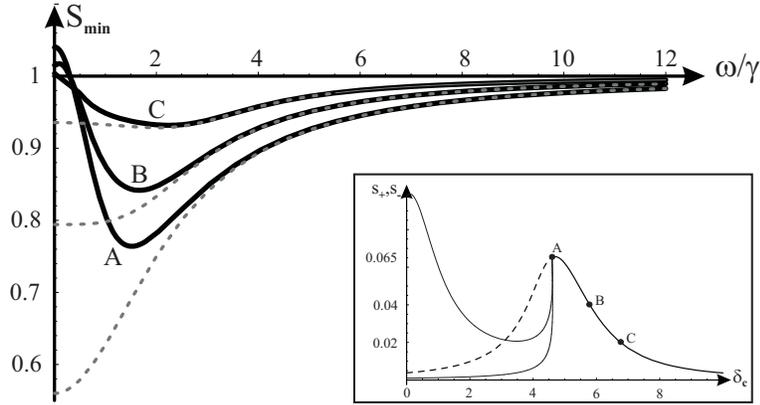}
  \caption{Minimal spectra for the vacuum field (plain) and the mean field (dashed) for different working points.
$\delta_0=5$, $\Delta=20\gamma$, $\kappa=2\gamma=6\gamma_{\perp}
$, $s_{max}=0.1$. The working points coordinates are: A
($\delta_c=\delta_{SR}=4.6$, $s_x/2=0.065$), B ($\delta_c=5.78$,
$s_x/2=0.04$), C ($\delta_c=6.79$, $s_x/2=0.02$). The inset shows
the working points positions on the resonance curve.}\label{fig8}
\end{figure}

As mentioned in Sec \ref{competition}, we expect squeezing to
improve in the vicinity of the PS threshold. We verified this
behavior by plotting on Fig \ref{fig8} the evolution of the
spectra when the cavity is scanned while keeping the incident
intensity ($\propto s_{max}$) constant. It appears clearly that
the best squeezing is obtained at the peak of the resonance curve,
right before the switching. This is due to the fact that, in the
low saturation regime, the PS threshold is close to the point
where saturation process is the most efficient.

\begin{figure}
  \centering
  \includegraphics[width=10cm]{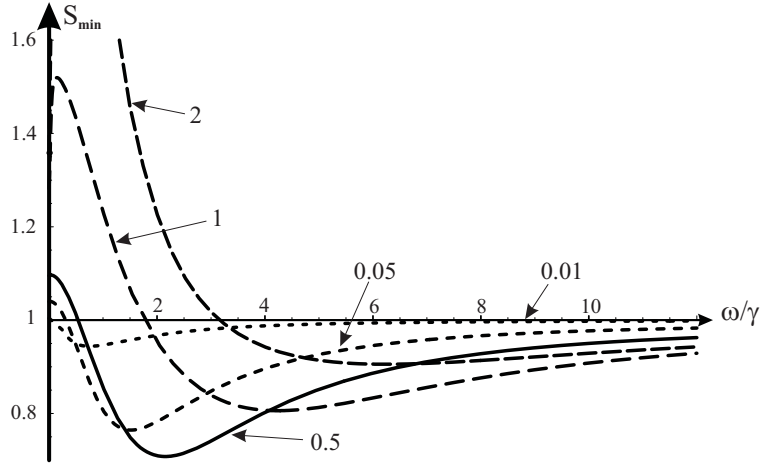}
  \caption{Minimal spectra for the vacuum field for different saturation values $s_{max}$.
  For each value of $s_{max}$, the working point is chosen close to
  PS. Parameters: $\delta_0=5$, $\Delta=20\gamma$,
  $\kappa=2\gamma=6\gamma_{\perp}$.}\label{fig9}
\end{figure}

\begin{figure}
  \centering
  \includegraphics[width=10cm]{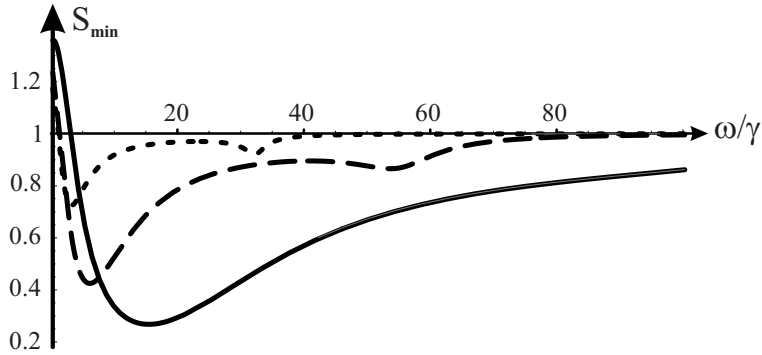}
  \caption{Minimal spectra for the vacuum for different $\rho=\kappa/\gamma$ (short dash: $\rho=2$,
  long dash: $\rho=10$, plain: $\rho=50$). For each value of $\rho$, the saturation is
  optimized. Parameters: $\delta_0=5$, $\Delta=20\gamma$, $\gamma/\gamma_{\perp}=3$.
}\label{fig10}
\end{figure}

We then study the effect of saturation and plot on Fig \ref{fig9}
various spectra corresponding to working points close to PS with
increasing saturation. The conclusion is that, for given values of
the detuning $\Delta$ and linear dephasing $\delta_0$, there is an
optimal value of $s_x$ for squeezing. This is due to the fact that
the range for which SR adds noise increases with the saturation
and eventually destroys Kerr-induced squeezing. The optimal
saturation value thus corresponds to a compromise between added
noise and Kerr squeezing in the intermediate frequency range.
Therefore, a bad cavity is preferable ($\kappa\gg\gamma$), since
Kerr-induced squeezing occurs on a frequency range given by
$\kappa$ and SR destroys squeezing for frequencies smaller than
$\gamma$. Spectra for different cavities are represented in Fig
\ref{fig10}. The case $\kappa=2\gamma$ corresponds to the
experimental situation, "close-to-bad cavity", the other curves to
increasingly bad cavities. Since SR is effective on a range
smaller and smaller compared to the cavity bandwidth, its effect
becomes negligible, and 75\% of squeezing can be obtained.\\
The conclusion is that very interesting squeezing values can be
reached in the bad cavity limit for both the mean field mode and
the orthogonal field mode. In the next section, we establish the
link between polarization squeezing and the vacuum squeezing
obtained in our system.

\section{Polarization squeezing}\label{polarizationsqueezing}

The noise of the mode with orthogonal polarization with respect to
the mean field is commonly referred to as polarization noise.
However, the study of the polarization state fluctuations requires
the introduction of the quantum Stokes operators
\cite{chirkin,korolkova,korolkova2}

\begin{eqnarray}
S_{0}(t)=A^{\dag}_{x}A_{x}+A^{\dag}_{y}A_{y}&\hspace{0.1cm},\hspace{1cm}&
S_{x}(t)=A^{\dag}_{x}A_{x}-A^{\dag}_{y}A_{y}\\
S_{y}(t)=A^{\dag}_{x}A_{y}+A^{\dag}_{y}A_{x}&\hspace{0.1cm},\hspace{1cm}&
S_{z}(t)=i(A^{\dag}_{y}A_{x}-A^{\dag}_{x}A_{y})
\end{eqnarray}

To be consistent with the definition of our slowly-varying
envelope operators $A_x$, $A_y$, these Stokes operators are
time-dependent and expressed in number of photons per second
\cite{fabre}. They obey the following commutation relationships

\begin{eqnarray}
\left[S_{0}(t),S_{i}(t')\right]=0\hspace{0.7cm}and\hspace{0.7cm}
\left[S_{i}(t),S_{j}(t')\right]= 2i\epsilon_{ijk}S_{k}\delta(t-t')
\end{eqnarray}

with $i,j,k=x,y,z$ and then the spectral noise densities of these
operators, defined by $\langle
S_i(\omega)S_i(\omega')\rangle=2\pi\delta(\omega+\omega')V_{S_i}(\omega)$,
satisfy uncertainty relations

\begin{equation}
V_{S_{i}}(\omega)V_{S_{j}}(\omega)\geq\epsilon_{ijk}|\langle{S_{k}\rangle}|^{2}\label{Heisenberg}
\end{equation}

The coherent polarization state correspond to the case where both
modes $A_x$ and $A_y$ are coherent states. Then the noise
densities of the Stokes parameters are constant and all equal to $
V_{S_{i}}(\omega)= \langle S_0 \rangle = |\langle A_x
\rangle|^2+|\langle A_y \rangle|^2$ for $i=0,x,y,z$. The so called
polarization squeezing is achieved if one or more of these
quantities (except $V_{S_0}$) is reduced below the coherent state
value

\begin{equation}
\frac { V_{S_i}(\omega)}{S_0}\leq 1 \hspace{1.5cm} (i=x,y,z)
\end{equation}

If the mean field is x-polarized, then $\langle S_0
\rangle=\langle S_x \rangle= |\langle A_x \rangle|^2 $ and
$\langle S_y \rangle=\langle S_z \rangle =0$. At first order in
noise fluctuations, $\delta S_y$ and $\delta S_z$ read

\begin{eqnarray}
\delta S_y & = & |\langle A_x \rangle |\left( \delta A^{\dag}_y
e^{i\theta _x}+\delta A_y e^{-i\theta
_x}\right)\;=|\langle A_x \rangle |\;\delta X_{A_y}(\theta_x) \\
\delta S_z & = & i|\langle A_x \rangle |\left(\delta A^{\dag}_y
e^{i\theta _x}-\delta A_y e^{-i\theta _x}\right)=|\langle A_x
\rangle |\; \delta X_{A_y}(\theta_x+\pi/2)
\end{eqnarray}

where $\theta_x$ is the phase of the mean field and
$X_{A_y}(\theta)= A^{\dag}_y e^{i\theta}+A_y e^{-i\theta}$ is the
quadrature with angle $\theta$ of the orthogonal mode. Therefore
the fluctuations of these two Stokes parameters are proportional
to the quadrature noise of $A_y$ and the polarization squeezing of
$S_y$ and $S_z$ is simply related to the vacuum squeezing that we
have studied in the previous sections. The physical meaning of
this result is clear: let us choose $\theta_x = 0$, then geometric
jitter on the polarization is due to the intensity fluctuations of
$A_y$ ($\propto \delta S_y$), whereas the fluctuations of the
ellipticity are caused by the phase fluctuations ($\propto \delta
S_z$). In the general case, the squeezed and antisqueezed Stokes
parameters are found to be

\begin{eqnarray}
S_{sq} & = & \cos(\theta_x - \theta_{sq} )S_y + \sin (\theta_x -
\theta_{sq} )S_z\\
S_{antisq} & = & \sin(\theta_x - \theta_{sq} )S_y - \cos (\theta_x
- \theta_{sq} )S_z
\end{eqnarray}

Note that, unlike \cite{grangier,polzik,bachor,korolkova3}, there
is no need to lock the phase-shift difference $\theta_x -
\theta_{sq}$, since it is automatically done in this system; this
property appears clearly in Eq (\ref{Aykerr}) where $\langle
A_x^2\rangle/|\langle A_x\rangle|^2=e^{2i\theta_x}$. Since the new
set of the Stokes parameters $S_0$, $S_x$, $S_{sq}$ and
$S_{antisq}$ still satisfy the relationships (\ref{Heisenberg}),
we obtain polarization squeezing in our system as soon as any
quadrature of the vacuum field $A_y$ is squeezed, and the results
of the previous sections can be applied to the squeezed Stokes
component.

\section{Conclusion}

We have presented a study of polarization switching in an X-like
4-level atoms ensemble illuminated by a linearly polarized light
in an optical cavity. PS has been traced to self-rotation and
simple criteria allow for a clear understanding of the switching
effects and the multistable behavior of the system. The steady
state analysis enables one to figure out the interesting working
points for squeezing.\\
In terms of squeezing the respective contributions of SR and
saturation have been investigated and compared to a full quantum
calculation. Since the propensity for squeezing of SR is cancelled
by atomic noise at low frequency, the squeezing originates from
Kerr effect. The mean field mode is squeezed via the usual
saturation effects, whereas the vacuum mode squeezing is induced
by the mean field via crossed Kerr effect. Both SR and crossed
Kerr effects can be dissociated in a bad cavity configuration,
thus allowing for high squeezing values. Last, this vacuum
squeezing is shown to be equivalent to squeezing one Stokes
operator.

\appendix
\section{}

Using the input-output theory notations
\cite{reynaud},\cite{hilico}, we give here the expressions of the
susceptibility matrix $[\chi(\omega)]$ and the correlation matrix
$[\sigma(\omega)]$ for field $A_y$. In the high frequency limit,
they resume to those derived in \cite{hilico} in the large
detuning limit

\begin{eqnarray}
\nonumber[\chi(\omega)]_{Kerr}=\frac{1}{2\Delta}
\left(\begin{array}{cc}
  1 & 0 \\
  0 & 1
\end{array}\right)
  &+\frac{1}{2\Delta^2}\left(
  \begin{array}{cc}
  i\gamma+\omega & 0 \\
  0 & -i\gamma-\omega\end{array}\right)\\
  & -\frac{g^2}{2\Delta^3}\left(
  \begin{array}{cc}
  2|A_x|^2 & \varepsilon A_x^2 \\
  \varepsilon A_x^{*2} & 2|A_x|^2\end{array}\right)
\end{eqnarray}

$\varepsilon=-1$ yields the susceptibility matrix for the vacuum
mode. To retrieve the matrix for $A_x$, $\varepsilon$ should be
taken equal to $+1$. This matrix corresponds to approximating the
atoms ensemble with a Kerr medium: the term of order 1 in
$1/\Delta$ is the linear dephasing, the second order matrix
represents dispersion and absorption and the third order term is
the non-linear dephasing corresponding to the Kerr effect. The
associated correlation matrix is

\begin{equation}
[\sigma(\omega)]_{Kerr}=\frac{\gamma}{\Delta^2}\left(\begin{array}{cc}
  1 & 0 \\
  0 & 0\end{array}\right)
\end{equation}

In the Kerr limit, the atomic noise comes only from the frequency
independent linear losses of the Kerr medium, which acts as a
beamsplitter for the field. Similar matrices can be derived for
field $A_x$ in agreement with (\ref{Ax1}). At low frequency,
however, the previous matrices have to be completed by

\begin{eqnarray}
[\chi(\omega)]_{SR}=-\frac{1}{2\Delta}\frac{\gamma_p}{\gamma_p-i\omega}\left(
\begin{array}{cc}
  1 & -A_x^2/|A_x|^2 \\
  -A_x^{*2}/|A_x|^2 & 1\end{array}\right)\label{chiy}
\end{eqnarray}
\begin{eqnarray}
\nonumber[\sigma(\omega)]_{SR} & = &
\frac{\gamma_p^2}{4\gamma_{\perp}(\gamma_p^2+\omega^2)}\left(\begin{array}{cc}
  1 & -A_x^2/|A_x|^2 \\
  -A_x^{*2}/|A_x|^2 &
  1\end{array}\right)\\
  & &
  +\frac{\gamma_p}{2\Delta(\gamma_p^2+\omega^2)}\left(\begin{array}{cc}
  -2\omega & \omega+i\gamma_p\\
  \omega-i\gamma_p & 0\label{sigmay}\end{array}\right)
\end{eqnarray}

For $\omega\ll\gamma_p$, the vacuum correlation matrix is
equivalent to

\begin{equation}
[\sigma(\omega)]_{SR}\sim\frac{\gamma_p^2}{4\gamma_{\perp}(\gamma_p^2+\omega^2)}\left(\begin{array}{cc}
  1 & -A_x^2/|A_x|^2 \\
  -A_x^{*2}/|A_x|^2 & 1\end{array}\right)
\end{equation}

so that a lot of noise is reported on all the quadratures of $A_y$
for frequencies of the order of $\gamma_p$, as pointed out in Sec
\ref{faradaysection}. For frequencies $\omega\gg\gamma$, the SR
noise terms vanish, allowing for crossed Kerr effect to produce
squeezing.

\end{document}